\documentclass{jpp}

\usepackage{graphicx}
\usepackage{natbib}
\usepackage[colorlinks,urlcolor=blue,citecolor=blue,linkcolor=blue]{hyperref}
\usepackage{amsmath}
\usepackage{subcaption}
\usepackage[export]{adjustbox}
\usepackage{gensymb}

\ifCUPmtlplainloaded \else
  \checkfont{eurm10}
  \iffontfound
    \IfFileExists{upmath.sty}
      {\typeout{^^JFound AMS Euler Roman fonts on the system,
                   using the 'upmath' package.^^J}%
       \usepackage{upmath}}
      {\typeout{^^JFound AMS Euler Roman fonts on the system, but you
                   dont seem to have the}%
       \typeout{'upmath' package installed. JPP.cls can take advantage
                 of these fonts, if you use 'upmath' package.^^J}%
      }
  \else
  \fi
\fi

\ifCUPmtlplainloaded \else
  \checkfont{msam10}
  \iffontfound
    \IfFileExists{amssymb.sty}
      {\typeout{^^JFound AMS Symbol fonts on the system, using the
                'amssymb' package.^^J}%
       \usepackage{amssymb}%
         
         \let\geq=\geqslant
      }{}
  \fi
\fi

\ifCUPmtlplainloaded \else
  \IfFileExists{amsbsy.sty}
    {\typeout{^^JFound the 'amsbsy' package on the system, using it.^^J}%
     \usepackage{amsbsy}}
    {\providecommand\boldsymbol[1]{\mbox{\boldmath $##1$}}}
\fi

\newcommand{\be}{\begin{equation}}
\newcommand{\ee}{\end{equation}}
\newcommand{\nn}{\mbox{} \nonumber \\ \mbox{} }
\newcommand{\ba}{\begin{eqnarray}}
\newcommand{\ea}{\end{eqnarray}}

\newcommand{\curl}{{\rm curl\, }}

\newcommand{\B}{{\bf B}}
\newcommand{\J}{{\bf J}}

\renewcommand{\div}{{\rm \,div\,}}

\newcommand{\Bf}{{magnetic field}}

\title{Nonlinear  force-free configurations in cylindrical geometry}
\author{ Maxim Lyutikov\aff{1}
 \corresp{\email{lyutikov@purdue.edu}}}

\affiliation{\aff{1}Department of Physics and Astronomy, Purdue University, 525 Northwestern Avenue, West Lafayette, IN, 47907-2036, USA}

\begin{document}

\maketitle

\begin{abstract}
We find   a new family of solutions for   force-free magnetic  structures in cylindrical geometry. These solutions
have radial power-law dependance and are periodic but non-harmonic in azimuthal direction; they  generalize the vacuum $z$-independent  potential fields
to current-carrying configurations.
\end{abstract}

\section{Introduction}

Force-free magnetic configurations, satisfying condition $\B = \kappa \J$, where $\B$ is \Bf\  and $\J$ is current density,  are examples of magnetic structures that may  represent the final stages of magnetic relaxation, or can be used as building block of plasma models  \citep{Lundquist50,Woltier:1958,1974PhRvL..33.1139T,2000mare.book.....P}. 

Particular {\it linear}  examples of force-free equilibria, with  spatially constant $\kappa$,  were considered by  \cite{ChandrasekharKendall57}.  The most often-used configurations are Lundquist fields in cylindrical geometry  \citep{Lundquist50}  and spheromaks in spherical geometry \citep{BellanSpheromak}.

Using the  self-similar assumption \cite{1994MNRAS.267..146L} found non-linear self-similar solutions in spherical geometry. Their model of  axially symmetric twisted configurations has been widely used  in astrophysical and space applications \citep[\eg][]{2002ApJ...574..332T,2011LRSP....8....6S}.
In the spirit of \cite{1994MNRAS.267..146L}, 
in this paper we construct similar non-linear magnetic configurations  in cylindrical geometry.

\section{Self-similar configuration in cylindrical geometry}

\cite{Shafranov} and \cite{1967PhFl...10..137G} formulated what is known as the Grad-Shafranov equation, separating complicated magnetic configuration
in the set of nested/foliated flux surfaces, given by the condition that flux function  $P$ is constant on the surface, and the    encompassed current flow. 
Let us look for force-free equilibria that are independent of coordinate $z$. The two Euler potentials $\alpha$ and $\beta$ (or, equivalently,  the related Clebsh variables) are
\ba &&
\alpha = z
\nn &&
\beta = P(r,\phi)
\ea
while the \Bf\ can be written as
\be
\B=\nabla P \times  \nabla z +  g(P)  \nabla z 
\ee
where $g$ is some function.
 
 Next we introduce a self-similar anzats 
\ba &&
P(r,\phi) =  r^{-l}  f(\phi)
\nn &&
g(P) =  {\cal C}  |P|^p
\nn &&
\B = \nabla P \times  \nabla z + {\cal C}  |P|^p  \nabla z =  \{  f', l f,  {\cal C} |f|^{p} \}r^{-(l+1)}
\label{P}
\ea
 Absolute value of $|P|$ in the non-linear term ensures that \Bf\ is real ($f(\phi)$ can become negative). Below any appearance of $f$ to non-integer power is to be understood to involve $\sqrt{f^2}=|f|$.

By dimensionality
\be
p = 1+1/l
\ee
Equation for $f$ becomes
\be
l f^{\prime \prime} + l^3 f + {\cal C}^2 (1+l) f^{(2+l)/l}=0
\label{111}
\ee
(note that the component $B_z$ enters here as $B_z^2$. This justifies the use of $|P|$.) 

For vacuum fields $ {\cal C}=0$ the above relations reproduce
\ba && 
P\propto r^{-m} \sin ( m \phi) 
\nn &&
B_r \propto r^{-(1+m)} f'
\nn &&
B_\phi \propto m r^{-(1+m)} f
\label{vac}
\ea
with integer $m$.

 The  first integral is
\be
f^{\prime,2}  + l^2 f^2 + {\cal C}^2   |f|^{2 (1+l)/l} = H_0,
\label{H0}
\ee 
By redefining $f\rightarrow \sqrt{H_0} f $ and $ {\cal C} \rightarrow  {\cal C} H_0^{-1/(2l)}$ the parameter $H_0$ can be set to unity,
\be
f^{\prime,2}  + l^2 f^2 + {\cal C}^2   |f|^{2 (1+l)/l} =1.
\label{H01}
\ee
Equation (\ref{H01}) is the main equation describing non-linear force-free structures in cylindrical geometry.  It depends on one parameter - the current strength ${\cal C}$. For a given ${\cal C}$ the value of $l$ is then determined as an eigenvalue problem by requiring periodicity in $\phi$, as we describe next. 

We can  solve for $f$ in quadratures:
\be
\phi = \int   \left( \sqrt{1  - l^2 f -  {\cal C}^2   |f|^{2 (1+l)/l}} \right) ^{-1} df
\ee
(so that the integration constant in Eq. (\ref{H01}) is just a phase $\phi$ where $f=0$).

Periodicity in $\phi$ requires
\be
\int _0 ^{f_{max}}  \left( \sqrt{1 - l^2 f -  {\cal C}^2   |f|^{2 (1+l)/l}} \right) ^{-1} df = \frac{\pi}{2m}
\label{mani}
\ee
where $m=1,2..$ is an integer azimuthal number (see a comment after Eq. (\ref{modd}) why odd solutions, $\propto 2m+1$ in the denominator, are discarded),
The value of   $f_{max}$ satisfies
\be
 1  - m^2 f _{max}-  {\cal C}^2 f_{max}^{2 (1+l)/l} =0 
 \label{mani1}
\ee

For given  $  {\cal C}$ the  relations (\ref{mani})- (\ref{mani1}) constitute an eigenvalue problem on  $l$.  (For vacuum no current case ${\cal C}=0$ this reduces to $l=m$,  an  integer - checkmark.) In practice, 
we follow the following procedure: for each $m=1,2..$ we assume some $l$ and find ${\cal C}$ using relations (\ref{mani})- (\ref{mani1}).
Thus, for each $m$ there is a continuous relation   ${\cal C}(l)$. (Physically, of course, it is the current ${\cal C}$ the determines the radial index $l$.)

\begin{figure}
\includegraphics[width=.5\textwidth]{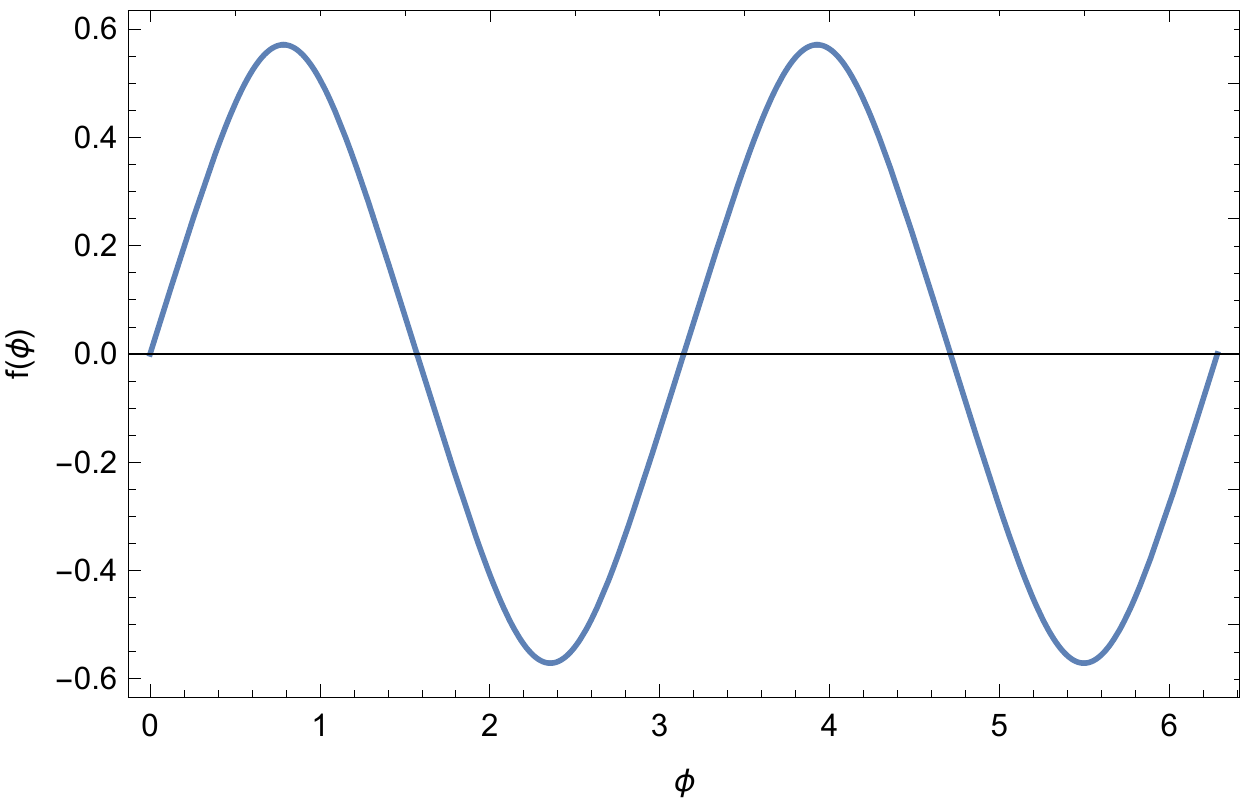}
\includegraphics[width=.5\textwidth]{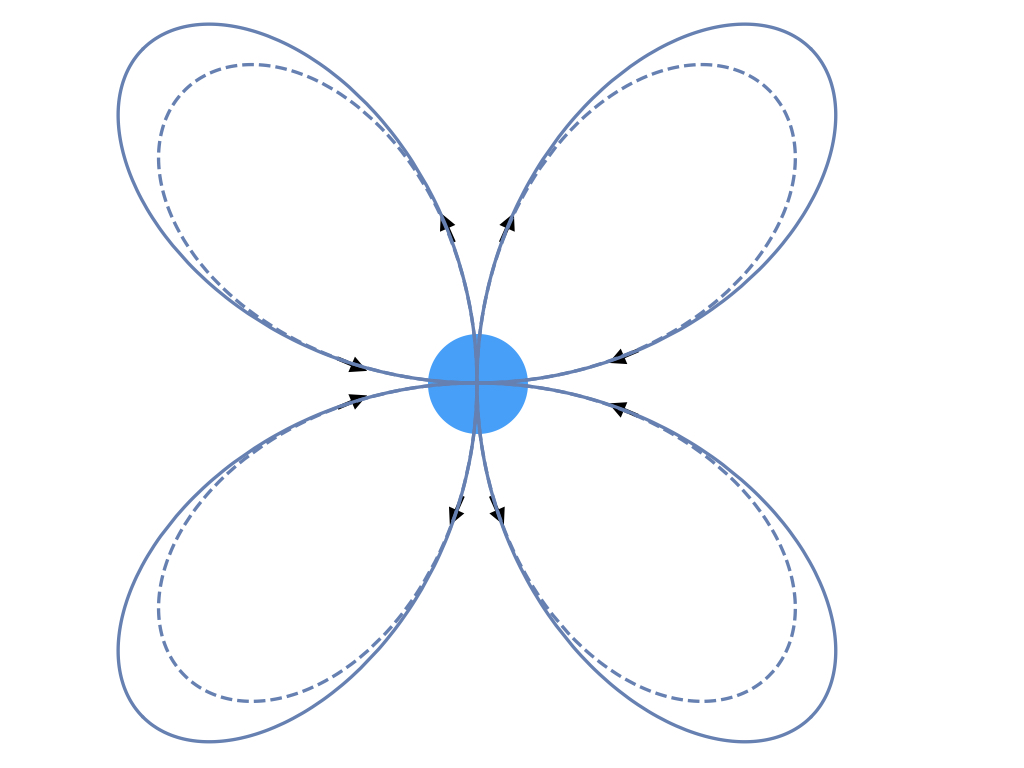}
\caption{Left panel: example of solution $f(\phi)$ for $l=1$, $m=2$. In this case ${\cal C}=2.517$.  Right panel: structure of poloidal field. (Due to the assumed self-similar radial structure the solutions do not extend to $r=0$). Dashed line is the corresponding vacuum case, Eq.  \protect\ref{vac}.}
\label{fofphi}
\end{figure}

\begin{figure}
\includegraphics[width=.99\textwidth]{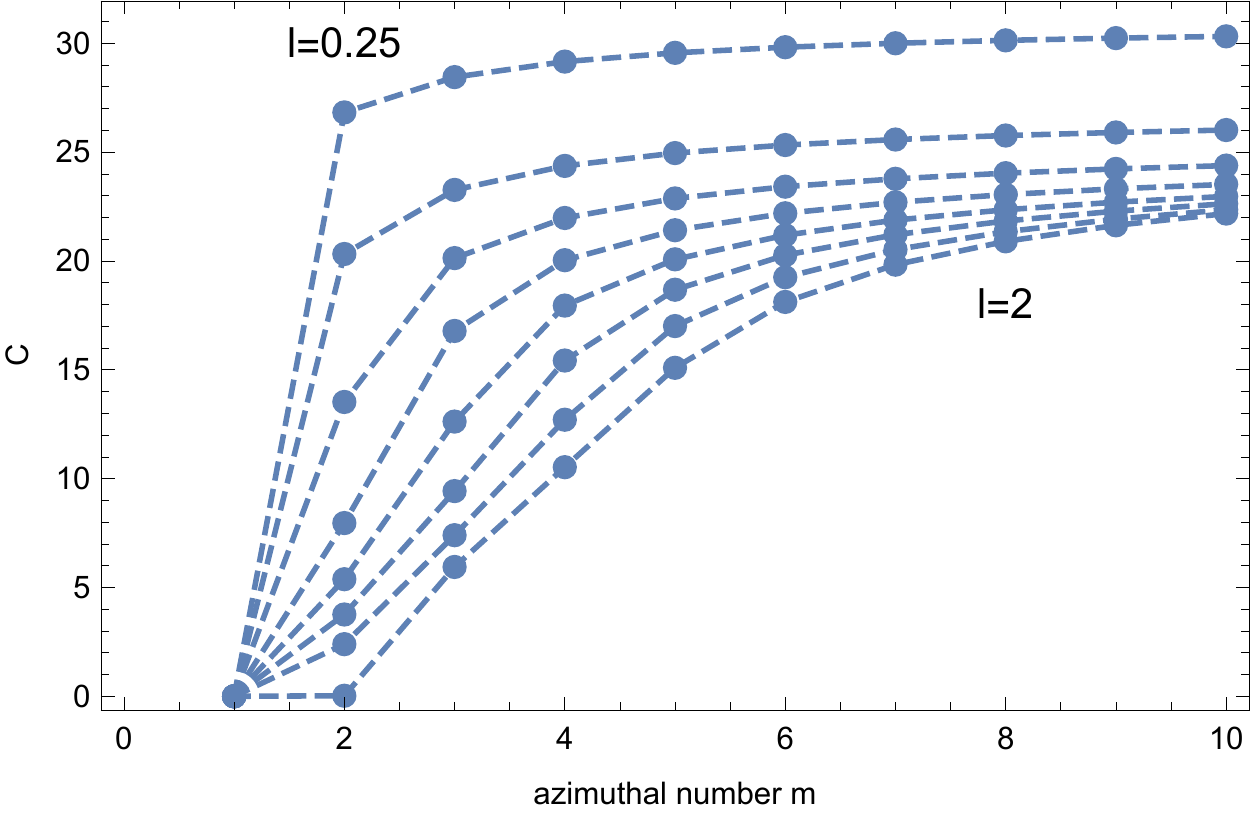}
\caption{Values of $\cal{C}$ as a function of azimuthal number $m$ for different values of $l=0.25...2$ (in steps of $0.25$). Top curves correspond to smaller $l$. }
\label{CofnH} 
\end{figure}

Results are plotted in Figs \ref{fofphi}-\ref{CofnH}. In Fig.  \ref{fofphi} we plot a particular solution for  $l=1$ and $m=2$. The flux functions forms a ''petal'' patter in azimuthal angle with number of ''petals'' equal $2m$. There is a corresponding axial, unidirectional \Bf\ $B_z$. 

\begin{figure}
\includegraphics[width=.99\textwidth]{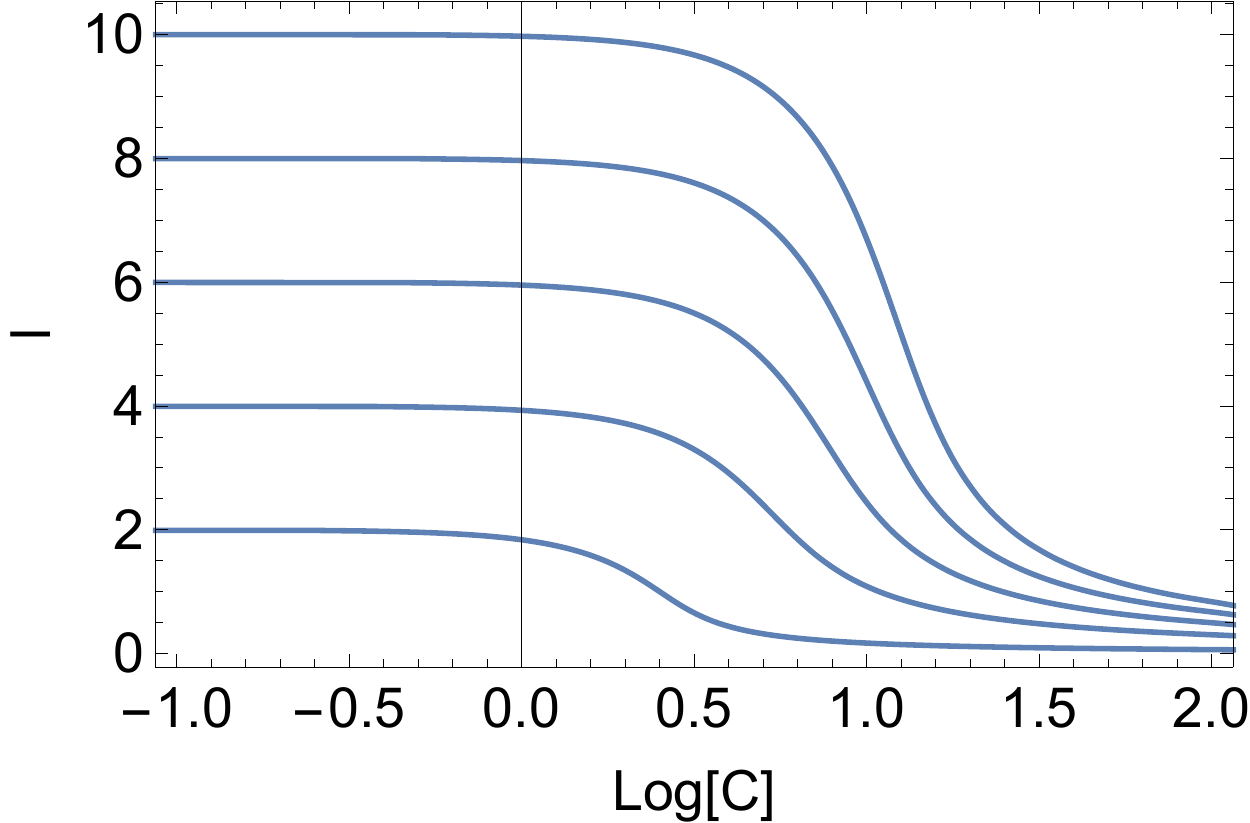}
\caption{Dependence  of the radial index $l$ on the current parameter  ${\cal C}$ for various harmonics $m=2,4,6,8,10$.}
\label{lofC} 
\end{figure}

In Fig. \ref{lofC}  we plot the curves $l({\cal{C}})$ for various $m=2,4,6,8,10$. Each curve starts at a point $\{  {\cal{C}} =0, l = 2m \}$. For non-zero current ${\cal{C}} >0$ the radial dependence becomes more shallow, $l< 2m$.

In Fig. \ref{CofnH} we plot values of $\cal{C}$ as a function of azimuthal number $m$ for different values of $l=0.25...2$. Dashed lines are for convenience only, they connect points corresponding to the same radial parameter $l$.

\section{Analysis of the solutions}

In a formulation of force-free  fields  in  the   form 
\be
\curl \B = \kappa \B
\ee
the value of $\kappa$ is
\be
\kappa =  {\cal C} \frac{1+l}{l r} f^{1/l}
\ee
It is constant on flux surfaces $P$, Eq. \ref{P}.

The current density (we incorporate factors of $4\pi/c$ into definition of \Bf)
\ba && 
j_r = {\cal C} \frac{1+ll}{l} r^{-(l+2)}  \partial_\phi \left( |f| \right) ^{(1+l)/l}
\nn &&
j_\phi=  {\cal C} (l+1) r^{-(l+2)}\left( |f| \right) ^{(1+l)/l}
\nn &&
j_\phi =  r^{-(l+2)} \left( f^{\prime \prime} + l^2 \right) =
- {\cal C}^2  \frac{1+ll}{l} r^{-(l+2)} \left( |f| \right) ^{(2+l)/l}
\ea

The total axial current is 
\be
I_z=\int_{r_0} ^\infty r dr \int_0 ^{2\pi}  d \phi  j_z=  - \frac{r_0^{-l}} {l}   \int_0^{2\pi} d\phi  \left(  f^{\prime \prime} +l^2  f\right).
\ee
where $r_0$ is the inner boundary. The  total axial current
vanishes if two conditions are satisfied 
\ba && 
\int  _0^{2\pi} f   d \phi  =0
\nn &&
f^{\prime} (2\pi) = f^{\prime}(0)
\label{modd}
\ea
All the solutions considered here satisfy these conditions: the second one requires even azimuthal numbers, $2m$.  Generally, there is a larger family of self-similar force-free equilibria with non-zero total axial current.

There is non-zero toroidal  current
\ba &&
j_\phi = {\cal C} (l+1) r^{-2 -l} |f|^{(1+l)/l}
\nn &&
\int _0 ^{2\pi} d \phi j_\phi \neq 0
\ea

The radial current density integrated over $\phi$ satisfies
\be
j_r \propto  \int_0 ^{2\pi}   d \phi \partial_\phi \left( |f| \right) ^{(1+l)/l}= \left. \left( |f| \right) ^{(1+l)/l} \right|_{0} ^{2\pi} =0
\ee

\section{Discussion}
In this paper we {\it make analytical progress with the highly nonlinear problem[s] of magnetohydrodynamics}  \citep{1994MNRAS.267..146L}.
We find a class of non-linear  self-similar force-free equilibria in cylindrical geometry.  The solutions we find all  connect to the vacuum case, in which case the  flux function is   $P_{vac} \propto 
r^{-m}  \sin ( m \phi)$.  Structures with vanishing total axial current require even values of $m$ (hence $m \rightarrow 2m $).
For non-zero distributed current  with the current parameter ${\cal C}$ the radial dependence changes to $r^{-l}$,  with $l< 2m$, while remaining periodic in $\phi$ at $2m$. Solutions for a given $m$ resemble vacuum solutions $\propto \sin ( 2 m \phi)$, but they are not exactly harmonic in the nonlinear case.

For very large currents the solutions asymptote to $l\approx 0$, but never reach this limit. The case $l=0$ corresponds to $B_r \propto 1/r$. Mathematically,   this is the analogue of split monopole case in the spherical geometry - split monopole case can be achieved in spherical geometry (with corresponding anti-monopole in the opposite hemisphere), but is not possible  in the cylindrical geometry.

\section*{Acknowledgments}
This work had been supported by 
NASA grant 80NSSC17K0757 and  NSF grants 10001562 and 10001521. 

\bibliographystyle{jpp}
\bibliography{/Users/maxim/Home/Research/BibTex}

\end{document}